# Study of Raspberry Pi 2 Quad-core Cortex-A7 CPU Cluster as a Mini Supercomputer

Abdurrachman Mappuji[1], Nazrul Effendy[2],
Muhamad Mustaghfirin[3], Fandy Sondok[4],
Rara Priska Yuniar[5]
Faculty of Engineering,
Universitas Gadjah Mada
Yogyakarta, Indonesia
[1]abdurrachman.mappuji@mail.ugm.ac.id,
[2]nazrul@ugm.ac.id

Sheptiani Putri Pangesti[6]
Faculty of Mathematics and Natural Science
Universitas Gadjah Mada
Yogyakarta, Indonesia
[6]sheptiani.putri.p@mail.ugm.ac.id

*Abstract*—High performance computing (HPC) devices is no longer exclusive for academic, R&D, or military purposes. The use of HPC device such as supercomputer now growing rapidly as some new area arise such as big data, and computer simulation. It makes the use of supercomputer more inclusive. Today's supercomputer has a huge computing power, but requires an enormous amount of energy to operate. In contrast a single board computer (SBC) such as Raspberry Pi has minimum computing power, but require a small amount of energy to operate, and as a bonus it is small and cheap. This paper covers the result of utilizing many Raspberry Pi 2 SBCs, a quad-core Cortex-A7 900 MHz, as a cluster to compensate its computing power. The high performance linpack (HPL) is used to benchmark the computing power, and a power meter with resolution 10mV / 10mA is used to measure the power consumption. The experiment shows that the increase of number of cores in every SBC member in a cluster is not giving significant increase in computing power. This experiment give a recommendation that 4 nodes is a maximum number of nodes for SBC cluster based on the characteristic of computing performance and power consumption.

## I. Introduction

Applications of supercomputer technology, both computer based vector processing, cluster computer, and commodity computer is growing rapidly, especially in data mining, which contributes in excavations information from the extensive data. Supercomputer technology also contributes in the simulation of many physical or life phenomena such as fluid dynamic computing, and multiphysics computing for the engineering purpose as well as for the development of science.

At least supercomputer is used for two main purposes i.e. the practical purpose and the development of science. Military purpose, data mining, weather forecast, web server, and video games are the example of the practical purposes of supercomputers. While the development of science, such as molecular simulation, life simulation, fluid dynamic computing, and many other scientific purposes which require a lot of computing needs are the second main purposes.

Data processing and data excavation are a main area of supercomputer applications today. The research, published by the Massachusetts Institute of Technology Sloan Management Review, shows that data mining can increase the income of the company until 187% [1] by taking advantage of the information was drawn from the market.

Nowadays, supercomputers require an enormous amount of energy to operate along with the increasing speed of supercomputers [2]. Even when in idle state (light up, but does not operate), it consumes a great power [3]. On the other side there are new recent research and development in single board computer (SBC). Summarizing from [4] today's SBC is ranging from a single-core CPU to a quad-core CPU, from 0.7 GHz CPU cycles/core to 2 GHz CPU cycles/core, without or with a Graphics Processing Unit (GPU), and ranging from $30 until $145. Some people have tried to make a cluster mini supercomputer with several of these SBC (they call it an embedded cluster computer, because the true main function of the SBC is for the embedded system problem), but there is still a common problem. SBC is a low power computer, but with minimal computing power, although there is some bonus such as small and cheap. It is usually powered by an ARM microprocessor compare to x86 or x64 model of microprocessor which is better on a number of core and the rate of CPU cycles.

Early research in embedded cluster computer or mini supercomputer has shown a proof of concept of the SBC's utilization as a cluster supercomputer [3][5]. However, it still shows a low computing performance of the cluster. A research conducted by Moore et al. [5] shows that a 4-nodes, single-core, 0.7 GHz cluster computer has 836 MFLOPS computing power comparable to Yellowstone Sandy Bridge Xeon 64 bit has 1.2 PFLOPS computing power. Another research conducted by Cloutier et al. [3] with the same SBC but with 32-nodes shows that it has 4.370 MFLOPS computing power. This research will aim to see the performance of the mini supercomputer if the CPU cycle is increased to 0.9 MHz and the CPU core is increased to 4 cores.

## II. METHODOLOGY

### A. System and Component Specification

Raspberry Pi 2 model B is chosen as the cluster member because the availability and technically well documented. It has an ARM Cortex A7 CPU with 0.9 GHz CPU cycles/core, quad-cores, one gigabyte of RAM, and Broadcom Video-Core IV 250 MHz GPU [6]. In total 8 of Raspberry Pi is used to form a cluster. Two times 5-port USB 2A power supply is used (10A full capacity). USB to µ-USB is used to connect power supply for Raspberry Pi. The D-Link Fast Ethernet Unmanaged Switch 16 Ports is used to make interconnection between the SBC. The Debian Linux is the SBC's operating system.

### B. Network Specification

A cluster computer needs a high speed bandwidth of data transmission. It is better to use a Gigabit Ethernet than Fast Ethernet, but a Raspberry Pi 2 Model B mini-computer doesn't have a Gigabit Ethernet or USB 3 port, it has only Fast Ethernet. In this work a fast Ethernet cables and switch used to connect all the mini-computer together. The interconnection utilizes mesh topology to provide inter SBCs communications.

### C. Inter SBCs Communications

Computer, in general, involve many circumstances in which one process needs to exchange information with another process. Generally, general purpose computer utilizing communications via shared memory, in which two or more processes read and write to a shared section of memory, or more processes read and write to a shared section of memory [7]. Communications via shared memory is unreliable because we have both CPU (sometimes many CPUs) and memory in every cluster member. In this case message passing is used to coordinate the cluster member, in which packets of information in predefined formats are moved between processes by the manager called interface. The standard message passing interface (MPI) is used because standardization, portability, performance opportunities, functionality and availability. The architecture model of the cluster is shown in the Fig. 1 below which involve many CPUs and memory in every cluster member [8].

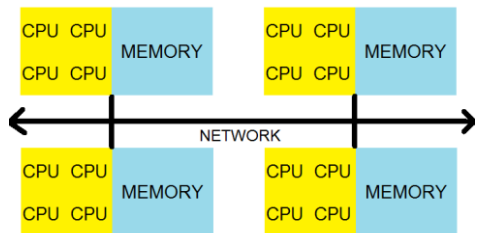

Fig. 1. Cluster architecture used in this paper work.

### D. Data Collection

In this work some related data are collected, there are power consumption vs. number of nodes, and computing performance (in MFLOPS) vs. number of nodes. Also in this work Raspberry Pi 2 Model B with quad-core 0.9 GHz CPU cycles are compared to overclocked old Raspberry Pi with single-core 0.9 GHz (controlled variable) CPU cycles. A voltage meter in a range 3.5V – 7.0V and a current meter in a range of 0A – 3A are used to measure and calculate the power consumption. The resolution of the power meter are 10mV / 10 mA and has operating temperature in a range of 0 – 60 °C. An example of measurement setup is shown in the Fig. 2. The power meter are plugged directly to every single power supply port and measure individual port power consumption. The data are shown every five seconds and the data are recorded manually.

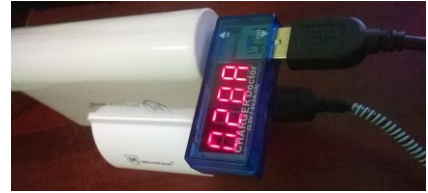

Fig. 2. The power meter measuring instrument.

### E. High Performance Linpack Benchmarking

In brief, HPL is a High-Performance Linpack benchmark implementation. The code solves a uniformly random system of linear equations and reports time and floating-point execution rate using a standard formula for operation count. HPL is written in a portable ANSI C and requires an MPI implementation as well as either BLAS or VSIPL library. Such choice of software dependencies gives HPL both portability and performance. HPL is often one of the first programs run on large computer installations to produce a result that can be submitted to TOP500 [9]. The HPL benchmark is used to benchmark the cluster performance in various numbers of nodes. Eight double precision words and 5040 matrix size is used while benchmark the cluster with the HPL.

### F. Design Optimization

The collected data are mixed and processed to gain the optimum node information. The power consumption is increased every computer-node added, but the computing power also increased every computer node-added. There is an optimal point we can gain. This optimization point one of these research goals.

## III. RESULT AND DISCUSSION

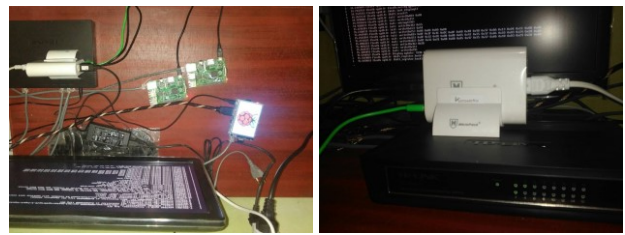

Fig. 3. Hardware for data collection

With the experimental setup and detail descripted in earlier section, this section provides the result and discussion of the experiment. The hardware for the data collection is shown in Fig. 3.

## A. Power Consumption vs. Number of Nodes

The graph shown below in Fig. 4 is the result of the power consumption measurement compared to the number of nodes utilized in the cluster. The raw data collected show that the working voltage of every individual SBC is constant in 5.16 V but the differences between the idle and working condition is the current utilizing. When in idle condition the current utilized is 0.23A in average while in working condition is 0.28A. So in average every single SBC is utilizing 1.19 W in idle condition and 1.44 in working condition.

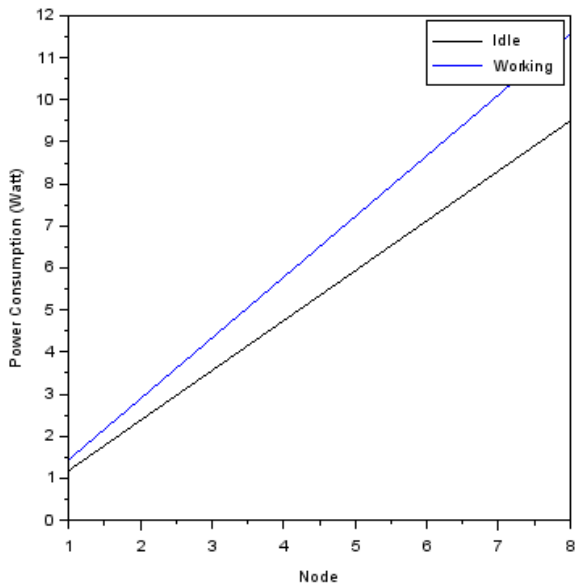

Fig. 4. Power consumption of the cluster in various node utilized both in idle and in working condition

## B. Computing Performance vs. Number of Node

Computing performance in MFLOPS or mega floating-point operations per second compare to number of node both in single-core and quad-cores are measured and shown in Fig. 5 and Fig. 6. Fig. 5 tells us that computing performance is not linear and tends to reach a saturation state as the node increased.

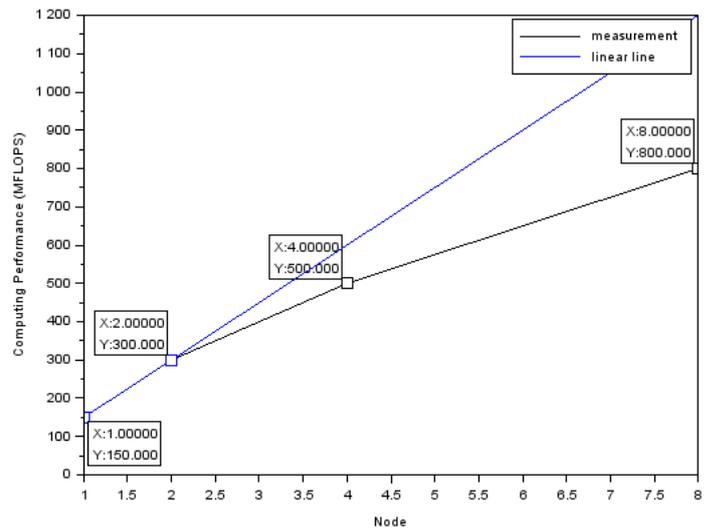

Fig. 5. Computing performance versus number of nodes in 0.9 GHz quad-cores Raspberry Pi 2 SBC compare to expected linear characteristic

Fig. 6 tells us that there is an improvement as core increased from 1 to 4 but we need to make a statistical test to make a statistically valid conclusion. A T-Test of two sample assuming unequal variance was conducted to test the effect of adding more cores in every member of SBC to whole cluster performance.

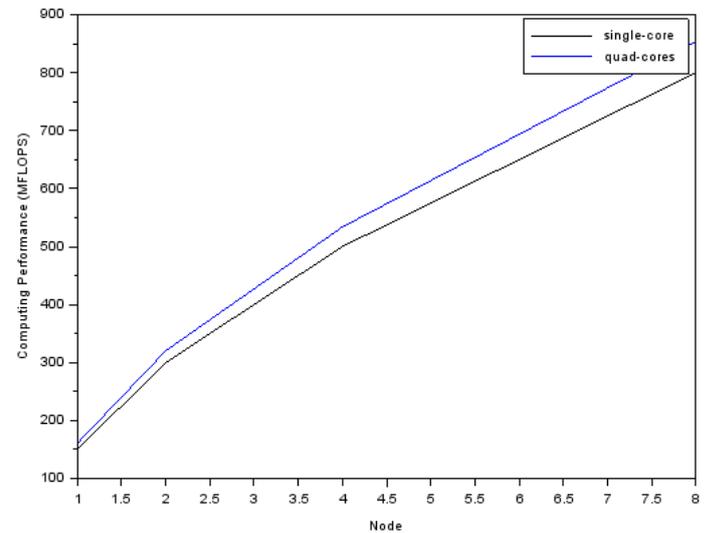

Fig. 6. Computing performance vs. number of nodes both in single core and quad-cores Raspberry Pi SBC

The t-Test is conducted with significance level (α) of 0.05, the null hypothesis ($\mu_0$) is mean performance of cluster with single-core and mean performance of cluster with quad-cores are the same, and the alternative hypothesis ($\mu_1$) is mean performance of cluster with single-core and mean performance of cluster with quad-cores is different. The test result is shown in Table 1. Because the test shows that the tStat parameter are less than tCritical one-tail, and P is greater than α, we conclude that we accept the $\mu_1$ and so that there is no significant difference between using single-core and quad-cores. It is not as the expected result because with the core increased it is expected that the computing performance is increased, but the result show the opposite. It may the result of the operating system resource management that not optimize to controlling multicore devices or the benchmarking program that not utilize concurrent programming concept like thread.

TABLE I. THE T-TEST CONDUCTED FOR SINGLE-CORE AND QUAD-CORE COMPARISON

|  | single core | quad cores |
|---|---|---|
| Mean | 437,5 | 467,1814306 |
| Variance | 78958,33333 | 89871,06118 |
| Observations | 4 | 4 |
| Hypothesized Mean Difference | 0 | |
| df | 6 | |
| t Stat | -0,144474365 | |
| P(T<=t) one-tail | 0,444928298 | |
| t Critical one-tail | 1,943180281 | |
| P(T<=t) two-tail | 0,889856595 | |
| t Critical two-tail | 2,446911851 | |

*C. Design Optimization Recommendation*

Design optimization recommendation result is generated by normalizing the computing performance vs. number of nodes graphic to be able to combine computing performance and power consumption graphic. The Fig. 7 is shown it combination. The picture below shows that when the node is below or equal two, the slope of the power consumption curve and computing power curve both has the same slope, but with the number of node is increased the slope of the computing power is decrease and make the addition of the number of node not worth compare to power consumption. With this data, it is recommended to use 4 nodes in maximal because the slope different is not big and still worth to add a node compare to power consumption. And also two nodes cluster is too small to make a powerful cluster computer.

IV. RECOMMENDATION AND FUTURE WORK

Because the parallelism of a multi-core computer, SBC is needed to optimize the computing performance. It is better to test the cluster with parallel optimized HPL benchmarking program with utilization of concurrent programming concept like thread and mutual exclusion in the future. Utilization of another kind of operating system is needed to enrich the knowledge of SBC cluster computer in the future work.

V. CONCLUSION

A research in utilizing multicore SBC as a cluster computer is conducted. The result shows that statistical test conducted prove that increasing number of cores in every member of SBC cluster is not giving significant increase in computing power. It may the result of the computer program that benchmark the program not utilizing potential performance of the multi-core device or resource manager of the operating system is not well suited for multi-core CPU.

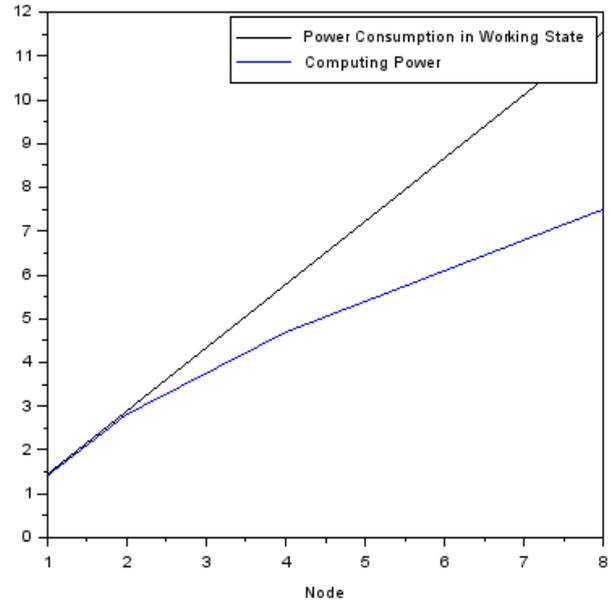

Fig. 7. Computing performance and power consuption graphic plotted in one graph by normalizing the computing performance vs. number of nodes to the factor of 7.5.

Beside that result, it is recommended to use maximum 4 nodes to form a SBC cluster because the increase of number of nodes, the decrease the slope of the computing performance curve and the increase of the computing performance is not worth compared with the power consumption.

ACKNOWLEDGMENT

The authors acknowledge the Ministry of Research, Technology, and Higher Education of Republic of Indonesia for supporting the research.